\documentclass[aps,pra,floatfix,showpacs,noshowkeys,epsfig]{revtex4-1}

\usepackage{graphicx}
\usepackage{amsfonts}
\usepackage{amssymb}

\newcommand{\newc}{\newcommand}
\newc{\beq}    {\begin{equation}}
\newc{\eeq}    {\end{equation}}
\newc{\PRD}  {{ Phys. Rev.} { D } }

\newc{\beqa}    {\begin{eqnarray}}
\newc{\eeqa}    {\end{eqnarray}}
\newc{\no}    {\\ \nonumber}

\def\PRL{{ Phys. Rev. Lett. }}

\def\PRD{{ Phys. Rev.} { D } }

\newc{\st}    {\stackrel}

\begin{document}
\title{ On the   Origin of  Entropic Gravity and Inertia }
\author{Jae-Weon Lee}
\address{ Department of energy resources development,
Jungwon
 University,  5 dongburi, Goesan-eup, Goesan-gun Chungbuk Korea
367-805}

\date{\today}

\begin{abstract}
It was recently  suggested
that  quantum field theory is  not fundamental but emerges from the loss of phase space
 information about matter
crossing causal horizons. Possible connections between
this formalism and  Verlinde's  entropic gravity and Jacobson's thermodynamic gravity
are proposed.
 The holographic screen in Verlinde's formalism can be identified as local Rindler horizons
  and its entropy as that of the bulk fields beyond the horizons.
    This naturally resolves some issues on entropic gravity.
  The quantum fluctuation of the fields is the origin of the thermodynamic nature of entropic gravity.
   It is also suggested that inertia is related to dragging
 Rindler horizons.
 \end{abstract}

\maketitle
\section{Introduction}


Studies of black hole physics have
 implied  a deep connection between gravity and thermodynamics~\cite{Bekenstein}.
 This leads to  Jacobson's proposal  that Einstein's
equation of general relativity can be derived from the first law of thermodynamics and the area-proportional Rindler horizon entropy~\cite{Jacobson}.
 Similarly, Verlinde~\cite{Verlinde:2010hp} recently
proposed  a remarkable new idea linking classical gravity to entropic forces,
which has attracted much interest
~\cite{Zhao:2010qw,Cai:2010sz,Cai:2010hk,Myung:2010jv,Liu:2010na,Tian:2010uy,Pesci:2010un,Diego:2010ju,
Vancea:2010vf,Konoplya:2010ak,Culetu:2010ua,Zhao:2010vt,Ghosh:2010hz,Munkhammar:2010rg,Kuang:2010gs}.
He derived  Newton's second law and  Einstein's equation from the relation
between the entropy of a holographic screen and mass inside the screen.
Padmanabhan ~\cite{Padmanabhan:2009kr} also  proposed that
 classical gravity can be derived from the equipartition law.
 However, the exact origin of this gravity-thermodynamics correspondence is unknown.

We proposed
a slightly different, but related  idea connecting  gravity to quantum information
~\cite{Lee:2010bg,Lee:2010fg} on the basis of Jacobson's model.
We suggested that  Einstein's equation can be derived from  Landauer's principle
by considering  information loss  at  causal horizons.
 Furthermore,
  it is also suggested \cite{myquantum} that  quantum mechanics
is not fundamental but emerges from  the information loss
and that this approach even can lead to the holographic principle~\cite{myholography}.
In this paper, it is suggested that the origin of   Verlinde's formalism and Jacobson's model
could be  explained by this information theoretic interpretation of quantum mechanics.

In quantum information science and gravity theories
 there are increasing number of perspectives that the
 universe is describable by information.
For example,
Zeilinger and Brukner
suggested that quantum randomness aries from the discreteness
of information~\cite{Zeilinger}.
't Hooft suggested that quantum mechanics
is a deterministic theory with local information loss~\cite{determinism}.
Our theory is in parallel with these ideas.

Inspired by the information theoretic nature of entropy,  in a series of works
  ~\cite{myDE,Kim:2007vx,Kim:2008re}
 we emphasized the  possible role of  information in gravity.
For example, in Ref. \cite{myDE}, we showed that a cosmic causal
horizon with a radius $R_h$ and Hawking temperature $T_h$ has a kind of thermal
energy $E_h\propto T_h S_h$ associated with holographic
 horizon entropy $S_h\propto R_h^2$ linked with the information loss at the horizon, and this thermal
energy (possibly the energy of the cosmic Hawking radiation) could be dark energy.
Using a similar approach the first law of black hole thermodynamics
is derived from the second
law of thermodynamics~\cite{Kim:2007vx}.
These works are based on the holographic principle
saying that the maximum number of independent degrees of freedom (DOF) in a region is proportional to its area
and   Landauer's principle.
This dark energy theory predicts the observed dark energy equation of state and magnitude~\cite{myDE}
 and the zero cosmological constant~\cite{Lee:2010ew}.
 (See ~\cite{Li:2010cj,Zhang:2010hi,Wei:2010ww, Easson:2010av, Easson:2010xf} for similar approaches
 based on Verlinde's formalism.)
 All these results imply that gravity has something to do with information loss at causal horizons.


Let us first briefly review Verlinde's formalism about Newton's classical mechanics.
In the formalism, inspired by the holographic principle, it was conjectured that  a
 holographic screen
with energy $E_h$, temperature $T$ and an area $A$ satisfies the equipartition law of energy
$E_h={k_B T N}/{2},$
where
$N=A/l_P^2$
 is the number of bits on the screen, $k_B$ is the Boltzmann constant, and $l_P$ is the Planck length.
 Verlinde argued that  a particle with mass $m$ approaching
the holographic screen  brings
the change of the entropy of the screen
\beq
\label{dSdx}
 \Delta S_h=\frac{2\pi c k_B m \Delta x}{\hbar},
\eeq
where $\Delta x$ is the distance between the screen and the particle, and $c$ is the light velocity.
It is one of the key assumptions of his paper allowing
the derivation of Newton's equation.
From the first law of thermodynamics, $\Delta S_h$ results in a variation of the screen energy
\beq
\label{1stlaw}
\Delta E_h=T \Delta S_h=F\Delta x,
\eeq
which can be interpreted to be the  definition of the holographic entropic force
$F\equiv \Delta E_h/\Delta x$.
If we use the Unruh temperature
\beq
\label{unruh}
 T_U=\frac{\hbar a}{2\pi c k_B}
 \eeq
 for $T$ and insert Eq. (\ref{dSdx}) into Eq. (\ref{1stlaw}),
 we  can immediately obtain the entropic force
 \beq
 \label{F1}
 F=\frac{\Delta E}{\Delta x}= T_U\frac{\Delta S}{\Delta x}=ma,
 \eeq
 which is just  Newton's second law.
Amazingly simple arguments reproduced one of the most basic equation in physics from thermodynamics.

However, there  appeared some criticisms~\cite{Li:2010cj,Culetu:2010ua,Gao:2010yy,Hossenfelder:2010ih} on
the  assumptions Verlinde took.
The most serious one seems to be about the origin of
the entropy and its variation equation in Eq. (\ref{dSdx}).
The usage of the second law of thermodynamics  apparently implies   the violation of the time reversal symmetry of gravity.
Furthermore, according to the holographic principle, the entropy increase should be proportional to some area
rather than the length scale $\Delta x$  as in Eq. (\ref{dSdx}).
It is also questionable whether we can apply the holographic principle to arbitrary surfaces
or  whether we can associate the Unruh temperature with the  surfaces.
The usage of the Unruh temperature  should be appropriately justified by careful analysis.
That is one of the aims of this paper.
The physical entity on the screen that has thermodynamic entropy is unclear.
Despite of the concerns proposed, considering remarkable implications of Verlinde's theory,
 it is  important to understand the exact physical origin of Verlinde's holographic screen and its entropy.

To overcome these difficulties, we reinterpret Verlinde's entropic force in terms of
the information theoretic model of quantum mechanics~\cite{myquantum} recently suggested.
This new interpretation shall turn out to be consistent with not only  Verlinde's formalism
but also
Jacobson's model~\cite{Jacobson} or quantum  information theoretic model of gravity~\cite{Lee:2010bg,Lee:2010fg}.
The key idea in this paper is that
for an accelerating object there is always an
 observer to whom the object seems to cross a local Rindler horizon of the observer.
 In this paper we assume the holographic principle and the metric nature of the spacetime (but not the Einstein equation).

 In Sec. II, the information theoretic interpretation  of quantum and classical mechanics  is
 briefly reviewed.
In Sec. III, the connection between this interpretation and  entropic gravity  is
presented.
Sec. IV contains discussions.

\section{mechanics from information loss}

In this section an information theoretic interpretation of the thermodynamic model for mechanics is presented.
To do this
let us first briefly review the information theoretic interpretation of quantum field theory (QFT) suggested in Ref. ~\cite{myquantum}.
Imagine an
accelerating observer $\Theta$ with acceleration $a$ in the $x_1$ direction
 in a flat spacetime with coordinates $x=(t,x_1,x_2,x_3)$ (See Fig. 1).
 There is another observer who is at rest.
The corresponding Rindler coordinates $\xi=(\eta,r,x_2,x_3)$  are
$t= r~ sinh (a \eta),~ x_1= r ~cosh (a \eta)$.
Now, consider a scalar field $\phi$ with Hamiltonian  $H(\phi)$,
which becomes  $H_R(\phi)$ in the Rindler coordinate.
.

As the field crosses  the Rindler horizon for the observer $\Theta$
and enters the future wedge $F$,
 the observer obtains no further information about  future configurations  of $\phi$,
 and
all that the observer can do at best is to estimate
  the probabilistic distribution $P[\phi]$ of $\phi$ beyond the horizon.
 The maximum ignorance about the field can be represented  by maximizing the Shannon
information entropy
$
h[P]=-\sum_{i=1}^n P[\phi_i] ln  P[\phi_i]
$ of  $n$ possible   field configurations
$\{\phi_i(x)\}~, i=1\cdots n$.

\begin{figure}[tpbh]
\includegraphics[width=0.3\textwidth]{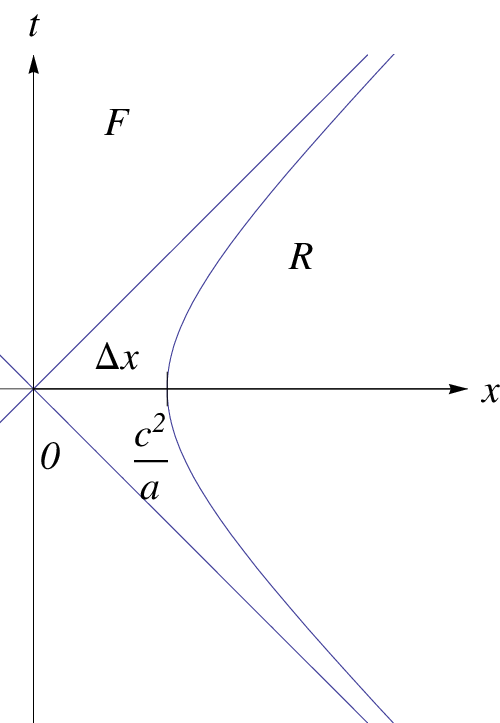}
\caption{
Rindler chart for  the observer $\Theta$.
The spacetime is divided into two causally disconnected  regions $F$ and $R$ by the Rindler horizon.
If $\Delta x=c^2/a$, the test particle is at the horizon for the observer
who would see the particle cross the local Rindler horizon.
}
\end{figure}

In Ref. ~\cite{myquantum} it was shown that
the required probability is
$P[\phi_i] = \frac{1}{Z_R} \exp\left[- \beta H_R(\phi_i)\right],$
where $\beta$ is a Lagrangian multiplier, and
 $Z_R$ is the partition function $ Z_R = \sum_{i=1}^n  \exp\left[- {\beta H_R(\phi_i)} \right]= tr~ e^{-\beta H_R}$.
 Lisi suggested a similar derivation of the partition function
by assuming a universal action reservoir for information~\cite{lisi-2006}.
Munkhammar also showed that quantum mechanics emerges from classical physics
with incomplete information~\cite{Munkhammar}.
Our  information theoretic
approache to quantum mechanics  reaches similar results.

The equivalence of $Z_R$
and  a quantum partition function for a  scalar field
in the Euclidean flat spacetime (say $Z^E_Q$)
 was shown by Unruh (the Unruh effect)~\cite{1984PhRvD..29.1656U}.
Thus, the conventional
  QFT formalism is equivalent to the purely  information theoretic formalism
  for loss of information about field configurations beyond the Rindler horizon,
  and the thermodynamic nature of QFT  naturally arises in this formalism.
  (See Ref. \cite{myquantum} for details.)
  The quantum fluctuation is actually thermal for the Rindler observer.

Let us derive the first law-like relation $dE_h=TdS_h$.
From $Z_R$ one can obtain various thermodynamic relations
between quantities such as $E$ (the total energy inside the horizon) or $S$ (the entropy related
to the phase space of matter inside the horizon, $h[P]$)
observed by the outside observer $\Theta$.
The   free energy $G=E-T_U S$ from the partition function is expressed as
$G=-\frac{1}{\beta} ln Z_R.$
Let us assume that spacetime is static and $T_U$ is constant.
Since the maximum entropy is achieved when $G$ is minimized (i.e., $dG=d(E-T_U S)=0$), we notice that
the maximum entropy condition due to the information loss, that has leaded  to quantum mechanics,
yields a condition
\beq
\label{1st}
dE=T_UdS
\eeq
for the fixed Unruh temperature.
This could be the physical
 origin of the relation, $dE_h=TdS_h$ used in Verlinde's formalism and Jacobson's formalism.
It  might also  explain the first law of thermodynamics in black hole physics~\cite{Kim:2007vx} and the cosmic expansion.

In our formalism,
the holographic entropy for entropic gravity is about the phase space
information loss of  the field beyond the horizon
rather than about
the exotic matter on the boundary like a string or a brane, and the maximum entropy condition is simply equivalent to
the  quantization condition.
The relation in Eq. (\ref{1st}) implies that small energy $dE$ of matter crossing the horizon induces an increase of the horizon entropy
\beq
\label{ds}
dS=dE/T_U
\eeq
by definition.
This equation  means that if we put the more matter into the horizon, the harder we can guess  the configurations inside the horizon.
The entropy and energy for the horizon are just those of the field inside the horizon.
In consideration of the definition of the horizon as an information blockade, this interpretation is  plausible.
We do not need to assume unidentified DOF on the horizon to explain the entropy, and
the thermodynamic nature of entropic gravity arises from the the very quantum nature
of QFT.

Shannon entropy represents lack of information about configurations
of a system \cite{Majumdar:1998xv}. Using this concept
Bekenstein calculated the black hole entropy by arguing that
the entropy is related to the equivalence class of all black holes having
the same external parameters such as  mass or charge~\cite{Bekenstein}.
Therefore, our information theoretic interpretation of $dS$ is  further justified by the black hole analogy.

Let us obtain the semiclassical limit of this system.
The classical field $\phi_{cl}$  corresponds to the saddle point that gives the largest contribution
to $Z_R$, and to the entropy. This also corresponds to extremizing the  Euclidean action  in $Z^E_Q$.
Then, the  semiclassical approximation using the Gaussian integral is
 $Z_R\simeq \frac{C}{\sqrt{\beta}} ~exp[-\beta H_R(\phi_{cl})]$,
 where $H_R(\phi_{cl})\equiv E_{cl}$ is the energy observed by $\Theta$
  for the classical field satisfying the Lagrange equation
 and $C$ is a constant.
In this limit the free energy  becomes
\beq
\label{semi}
G=-\frac{1}{\beta} ln Z_R\simeq -\frac{1}{\beta} \left\{-\beta H_R(\phi_{cl})\right\}-\frac{C'}{\beta}+O(\frac{ln\beta}{\beta})
=E_{cl} -\frac{C'}{\beta}+O(\frac{ln\beta}{\beta}),
\eeq
where $C'=ln~C$. Since $G=E-TS$, we immediately notice that $C'$ is a semiclassical entropy $S_{cl}$ seen by $\Theta$.
The minimum free energy condition $dG=0$ is now equivalent to
$dE_{cl}-T_U dS_{cl}=0$.
It explains why classical physics can be obtained from thermodynamics as in Verlinde's approach.
The classical field centers at the quantum field fluctuations which  maximize  the  entropy.
(See also \cite{Banerjee:2010yd}.)
This also explains why Verlinde's derivation involves $\hbar\sim 1/\beta$ which is absent in the final $F=ma$ formula.
If we identify $C'$ as the Bekenstein-Hawking entropy which is proportional to $1/\hbar\sim \beta$,
we can see $\hbar$ is canceled in the  term  $C'/\beta$.
The next order term is of $O(ln\beta / \beta)=O(\hbar~ ln \hbar)$, which is typical
in many semiclassical limit of quantum gravity models
and negligible in the classical limit $\beta\sim 1/\hbar\rightarrow \infty$.

Now, let us investigate the relation between this information theoretic theory and Verlinde's model.
It is straight forward to extend the above results for fields  to particle quantum mechanics.
 Since the conventional point particle quantum mechanics
 is a non-relativistic and single particle limit of QFT,
 we expect $Z_R$ for a particle with mass $m$  is equal to the quantum partition function for the particle
  in the Minkowski spacetime
 \beqa
 \label{Zq}
 Z_Q &=&
  N_2\int  Dx \exp\left[-\frac{i}{\hbar}\int d{t}  \left\{mc^2+\frac{m}{2}
 \left( \frac{\partial x}{\partial {t}}\right)^2 -  V(x)\right\}  \right],
\eeqa
 where we kept the leading rest mass term.
 Then,
as is well known one can associate each classical path $x({t})$ with
classical action $I(x)$ and a wave function $\psi\sim e^{-iI(x)}$,
which leads to  Schr\"{o}dinger equation for $\psi$.
Now, the partition function denotes the uncertainty of the path information
(i.e., phase space information) in this case.
A classical path is the typical path  maximizing the Shannon entropy $h[P[x]]$
of the paths seen by the Rindler observer.
Therefore,  the entropy associated with Newton's mechanics and the Einstein gravity is
related to the path information (i.e., phase space information) of particles beyond the Rindler horizon.
The information-energy relation  (Eq. (\ref{1st})) still holds, however, in  this case
 $E$ is the total mass of the particles inside the horizon and $S$ is the entropy associated with the paths of the particles.
Note that this identification is natural and logically consistent with the  conventional path integral formalism
invented by Feynmann.

\begin{figure}[tpbh]
\includegraphics[width=0.3\textwidth]{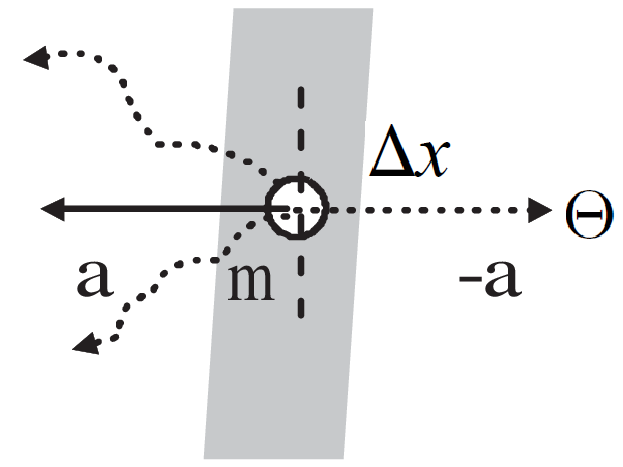}
\caption{
A test particle with mass $m$ is accelerating with acceleration $a$ with respect to
an observer $\Theta$ instantaneously  resting at $\Delta x$ from the particle.
Alternatively, according to the general principle of relativity, we can imagine that the particle is at rest
and the observer moves  in the opposite direction with acceleration $-a$.
The observer could see the particle crossing a Rindler horizon (the gray surface)
and the future paths of the particle (the dotted curves) become maximally uncertain to the outside observer.
This increases the horizon entropy.
}
\end{figure}

If this interpretation is consistent with  Verlinde's formalism, we should be able to reproduce Verlinde's entropy variation formula
(Eq. (\ref{dSdx})).
To check this  let us consider an accelerating  point particle  with acceleration $a$ and mass $m$ (Fig. 2)
and an  observer $\Theta$ at rest at the instantaneous distance $\Delta x=c^2/a$ from the particle.
The distance $\Delta x$
is special, because,  for the observer there, $\eta$ becomes a proper time and
 the Rindler Hamiltonian becomes a physical one
generating $\eta$ translation~\cite{1984PhRvD..29.1656U}.
If we accept the general principle of relativity stating that
all systems of reference are equivalent regardless of observer's motion,
 we can imagine an equivalent situation where the particle is at rest
and the observer $\Theta$ accelerates in the opposite direction with acceleration $-a$.
Then the observer can see the particle  cross the horizon.
 For this observer there should be information loss about the future paths of the particle.
The key idea here is that
for an accelerating object we can always assume  a Rindler observer to whom
   the object just appears to cross the Rindler horizon of the observer.
   This crossing is accompanied with increases of $S$ and $E$.

In the Newtonian limit(when $mc^2\gg m(\partial x/\partial {t})^2$), the horizon energy increase  is approximately the rest mass energy
$mc^2$ of the  particle, hence
 from  Eq. (\ref{1st}) and Eq. (\ref{unruh})  the following relations should hold.
\beqa
mc^2=\Delta E
=T_U \Delta S=\frac{\hbar a}{2\pi c k_B} \Delta S.
\eeqa
This equation means that if we put a particle  with mass $m$
 into the horizon, we are uncertain by $\Delta S$ about the paths that the particle may take.
For $a=c^2/\Delta x$ it straightforwardly gives Eq. (\ref{dSdx}) and Newton's equation.
(A similar derivation can be found in  Culetu's comments~\cite{Culetu:2010ua}.)
Since the entropy variation formula is successfully reproduced, we can identify  Verlinde's holographic screen  as
the Rindler horizon and the entropy of the screen $S$ as the entropy associated with the path seen by the  Rindler observer.

Considering $Z_Q$ in Eq. (\ref{Zq}), for a mildly relativistic case
 we can use an approximation $\Delta E\simeq mc^2+ {mv^2}/{2}$, where $v$ is the velocity of the particle.
Repeating the calculation with this $\Delta E$ gives a modified version of Newton's second law
\beq
F\simeq ma (1+\frac{v^2}{2c^2}),
\eeq
which is consistent with the usual formula for the  mildly  relativistic particles; $F\simeq \gamma ma$,
where $\gamma=(1-(v/c)^2)^{-1/2}$.

There are two ways to derive  classical mechanics in our formalism.
 First, conventional quantum mechanics based on $Z_Q$  of course reproduces classical Newtonian
  mechanics in  the  limit $\hbar \rightarrow 0$ and $c \rightarrow \infty$.
 Alternatively, one can also derive
 Newton's mechanics from the information-energy relation (Eq. (\ref{1st})) as shown above.
 One can easily check that both ways gives us the same classical physics  in the same limit,
 and hence,  our theory is consistent with the conventional mechanics.
 This fact further supports the information theoretic interpretation of entropic gravity.

Interestingly, this interpretation seems to also hint at the origin of inertia.
To understand it, let us return to the rest frame of the accelerating  particle.
In this frame the particle sees the Rindler horizon following  it at a distance $\Delta x$  and
can see part of its spacetime continually disappear behind the horizon. This induces the loss of information about vacuum quantum field.
(The thermal energy of the horizon is related to the vacuum fluctuation of quantum fields inside the horizon
and  was suggested to be dark energy~\cite{myDE}.)
Thus, to accelerate the particle or equivalently to
`drag' the horizon additional energy $\Delta E$ should be applied to the horizon.
This dragging requires  an external force $F=\Delta E/\Delta x$
exerted on the particle due to the action-reaction principle.
Thus, the inertia of the particle
can be interpreted as $resistance$ by the horizon that the external force feels.
This dragging force is proportional to acceleration, hence, $F=ma$.

 Haisch and Rueda~\cite{rueda-1998-240}   proposed a similar idea that
  inertia is from anisotropic distribution of vacuum
fluctuations due to the acceleration and even derived Newton's law.
In their model, however, inertia is  a consequence of  interaction with the electromagnetic zero-point field
and  depends on the specific interaction the matter has. On the other hand,
 in our model inertia is related to information loss and does not  depend much on
the interaction the matter has.

What happens if there is another particle?
In that case we need to consider the role of the second particle and
this  unavoidably  leads to  the law of universal gravitation described in the next section.

\section{Entropic gravity from information loss}

Let us reconsider Verlinde's derivation of the Newton's gravity in \cite{Verlinde:2010hp}.
Assume that a small test particle with mass $m$
is  at a distance $r$  from a massive object with mass $M$ at the center  as shown in Fig. 3.
The particle has acceleration $a$ due to the gravity of the object.
Verlinde conjectured that for this spherical system
 the energy of the screen with a radius $r$ is
 given by the equipartition rule and it is equal to the mass energy $M c^2$ inside the screen.
 Thus, the temperature $T$ is $2Mc^2/k_B N$.
Using the number of bits on the screen
$N={A}/{l_P^2}=4 S_{BH},$
and $A=4\pi r^2$ one can obtain
\beq
\label{T}
T=\frac{\hbar G M}{2\pi c k_B r^2},
\eeq
which is equal to the Unruh temperature for a gravitational acceleration.
Here,
$S_{BH}$ is the Bekenstein-Hawking entropy
\beq
\label{SBH}
S_{BH}=\frac{ c^3~A}{4G\hbar}=\frac{\pi r^2 c^3~}{G\hbar}
\eeq
which is an information bound a region of space with
a surface area $A$ can contain~\cite{Bekenstein:1993dz}.
 Inserting  Eq. (\ref{T}) and Eq. (\ref{dSdx})
 into Eq. (\ref{F1}) one can easily obtain
 the Newton's gravity formula
 \beq
 \label{F}
 F= T_U\frac{\Delta S}{\Delta x}=\frac{GMm}{r^2}.
 \eeq
 For strong gravity
 we need to consider  curved spacetime effects
 as in Ref. \cite{Jacobson,Lee:2010bg}.

\begin{figure}[tpbh]
\includegraphics[width=0.3\textwidth]{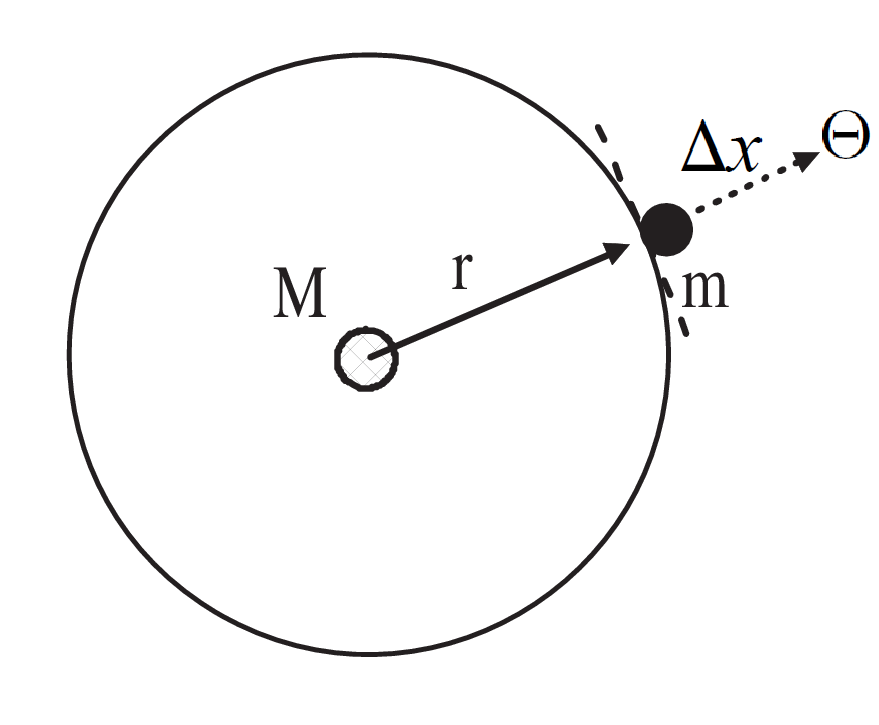}
\caption{
A test particle with mass $m$
is at a distance $r$  from a massive object with mass $M$ at the center.
Consider an accelerating  observer $\Theta$ with respect to the free falling frame
with acceleration $-a$.
If the observer is instantaneously at the distance $\Delta x=c^2/a$
from the test particle, the observer could see the particle crossing
the  local Rindler horizon (the dashed line) for the observer.
}
\end{figure}

Verlinde's simple derivation of Newton's gravity from thermodynamics is striking and full of suggestions, but
what is the origin of this thermodynamics?
 Similarly to the previous section,  we will reinterpret this derivation in terms of Rindler horizons.
The equivalence principle allows us to
 choose an approximately flat patch for each  spacetime point.
 According to the principle one can not locally distinguish the free falling frame
 from a rest frame without gravity.
We can again imagine an accelerating observer $\Theta$ with acceleration $-a$ respect to the test particle
in the instantaneous rest frame of the particle.
Note that this situation is similar to that considered by Jacobson to derive the Einstein equation~\cite{Jacobson}
from thermodynamics at local Rindler horizons.

If the observer is instantaneously at the distance $\Delta x=c^2/a$ from the test particle,
 the test particle is just crossing the   Rindler horizon for this specific observer $\Theta$.
As the test particle crosses the horizon  the mass of the  particle $m$
should be added to the horizon energy $E_h$, because the horizon energy in our formalism is simply the total energy inside the horizon.
This induces the increase of the horizon entropy $\Delta S_h$ by Eq. (\ref{ds}) due to the loss of the information about the paths of the test
particle.
Therefore, from Eq. (\ref{1st})
\beqa
mc^2=\Delta E_h
=T_U \Delta S_h=\frac{\hbar a}{2\pi c k_B} \Delta S_h
\eeqa
should hold.
Following the argument in the previous section
 one can obtain the entropy change in Eq. (\ref{dSdx}) again.

To derive Newton's gravity with our formalism
we still need to calculate $T_U$.
We can safely use the equipartition rule $E_h=N k_BT_h/2$
without assuming the specific thermal nature of the boundary physics, because the partition function
$Z_R$ represents a canonical ensemble. It is another merit of our formalism.
Here, $N$ is the number of DOF representing
the path information of the massive particle inside the horizon.
We also rely on the holographic principle.
Following Ref. ~\cite{Verlinde:2010hp} let us assume $N=4 S_{BH}$.
Therefore, the horizon energy is
\beq
Mc^2=E_h=\frac{N k_B T_U}{2}= 2 k_B T_U S_{BH}.
\eeq
From this equation one  obtains $T_U=Mc^2/2 k_B S_{BH}$  and
from Eq. (\ref{dSdx})   the entropic gravity force in Eq. (\ref{F}) arises.
This derivation is slightly different from that by Verlinde.
This information theoretic approach allows us to link Verlinde's formalism
to Jacobson's formalism.
Furthermore,  $E_h$ above is consistent with the
gravitational (Tolman-Komar) energy calculated by Padmanabhan \cite{Padmanabhan:2009kr}.
 He assumed the horizon entropy is related to the unobserved DOF beyond the horizon
as in our theory.

Here, for a mildly relativistic case
 we can  again use an approximation $\Delta E\simeq mc^2+ {mv^2}/{2}$, which
 gives a modified version of Newton's gravity law
 \beq
 \label{F}
 F=\frac{GMm}{r^2}(1+\frac{v^2}{2c^2}).
 \eeq

It is easy to extend this analysis to the Einstein gravity by following the derivation by  Jacobson.
We can  generalize the information-energy  relation $dE_h=TdS_h$  by defining the
energy flow across the horizon $\Sigma$
\beq
\label{generalized}
dE_h=-\kappa \lambda\int_\Sigma T_{\alpha\beta} \xi^\alpha d \Sigma^\beta,
\eeq
where $\kappa$ is the surface gravity, $\xi$ is a
boost Killing vector, $\lambda$ is an affine
parameter, $d\Sigma^\beta=\xi^\beta d\lambda dA$, $dA$ is a spatial area element,
and $T_{\alpha\beta}$ is the energy momentum tensor of  matter distribution.
(We set $c=1$.)
Using the Raychaudhuri equation one can
denote the horizon area expansion $\delta A\propto dS_h$  and the increase of the entropy
as
\beq
\label{dA}
dS_h=\eta' \delta A=-\eta' \lambda \int_\Sigma R_{\alpha\beta} \xi^\alpha  d\Sigma^\beta,
\eeq
with a constant  $\eta'=1/4 \hbar G$ ~\cite{Jacobson}.
By inserting Eqs. (\ref{generalized}) and (\ref{dA}) into Eq. (\ref{1st})
with the Unruh temperature $T_U=\hbar\kappa/2\pi$ one can find
$2\pi  T_{\alpha\beta} \xi^\alpha d \Sigma^\beta
=\hbar\eta'  R_{\alpha\beta} \xi^\alpha d \Sigma^\beta$. Due to the maximum entropy principle applied at the horizons,
for all local Rindler horizons this equation
should hold. Then, this condition and Bianchi indentity lead to
the Einstein equation
\beq
\label{einstein}
R_{\alpha\beta}-\frac{R g_{\alpha\beta}}{2}+\Lambda g_{\alpha\beta}
={8\pi G } T_{\alpha\beta }
\eeq
with the cosmological constant $\Lambda$~\cite{Jacobson}.

Therefore, we conclude that
the Einstein equation simply describes the loss of information  about matter crossing
local Rindler horizons in a curved spacetime.
The holographic screens  in Verlinde's formalism are
 actually local Rindler horizons  for accelerating observers relative to the test particle.
Albeit simple, this identification could easily resolve some questions of Verlinde's formalism
 and provide better grounds for the theory.
 The entropy-distance relation (Eq. (\ref{dSdx})) is naturally derived,
 and the use of  Unruh temperature is  justified, because the holographic screen is
 actually a set of local Rindler horizons.

However, there are  also several distinctions between Verlinde's  original model  and
our information theoretic interpretation of the model, which help us to resolve the possible difficulties
 of entropic gravity~\cite{Gao:2010fw}.
First,
in our theory,  spacetime is not necessarily emergent and the particle
just crosses the horizon rather than merges with it.
Second, in our theory the entropy change  happens only after the particle crosses the horizon.
This allows a more straight derivation of the entropy-distance relation
and helps us to avoid the purported problem associated with ultra-cold neutrons in the gravitational field~\cite{Kobakhidze:2010mn};
Kobakhidze argued that in Verlinde's model the entropy change in Eq. (\ref{dSdx}) tells us that the evolution of
the particle approaching the screen is non-unitary and this is inconsistent with the neutron experiment.
On the contrary, in our theory the entropy of the screen increases only when the particle crosses the horizon,
and in the rest frame of the particle this process is unitary, thus
the problem does not occur.
Third, in Verlinde's theory the
holographic screens correspond to equipotential surfaces, while
in our theory they correspond to isothermal Rindler horizons
(i.e., with the same $|a|$).
Finally, since the Rindler horizons are observer dependent, there is no
observer-independent notion of the horizon entropy increase in our theory.
Therefore, we do not need to worry about the issue of the time reversal symmetry breaking in entropic gravity.

\section{Discussions}
It was shown in this paper that the information theoretic interpretation of quantum mechanics
can explain the origin of entropic and Einstein's gravity.
If gravity is really entropic, identifying the underlying  microscopic DOF
  is very important.
We conclude that the holographic screens in Verlinde's formalism are
 actually local Rindler horizons  for specific  observers accelerating relative to the test particle,
 and the holographic entropy is from quantum fluctuations.
 This identification could resolve some issues on entropic gravity
 and provide a more concrete ground for it.
 The entropy-distance relation is  reproduced,
 and the use of  the Unruh temperature is well justified in this formalism.
 It also shows the interesting connection between
 Jacobson's model~\cite{Jacobson} or the  quantum information theoretic model~\cite{Lee:2010bg,Lee:2010fg}
 and Verlinde's model.
Our information theoretic approach supports both of Verlinde's  formalism and Jacobson's formalism  but with
the   modified  interpretation.

Alternatively, one can  reproduce the results
in this paper by simply taking the Unruh effect   and the holographic principle
as starting postulates. This might look more familiar for some readers.
However, in this case the meaning of the horizon entropy and the role of the holographic screen are
less obvious than the information theoretic interpretation.
%

Compared to the previous works by others, our theory emphasizes the role of information rather than thermodynamics.
In a series of work including this paper we proposed that information loss at causal horizons
is the key to understanding the origin of quantum mechanics, Einstein gravity, and even the holographic principle.
Another merit of the information theoretic approach is that it seems to provide us an explanation on the strange connections among
these different fields of physics~\cite{Lee:2010xa}.

\section*{acknowledgments}
Author is thankful to H. Culetu, Jungjai Lee, Hyeong-Chan Kim, and Gungwon Kang for
helpful discussions.
This work was supported in part by Basic Science Research Program through the
National Research Foundation of Korea (NRF) funded by the ministry of Education, Science and Technology
(2010-0024761) and the topical research program (2010-T-1) of Asia Pacific Center for Theoretical
Physics.


\end{document}